\documentclass[final,5p,twocolumn]{elsarticle}
\makeatletter
\def\ps@pprintTitle{%
 \let\@oddhead\@empty
 \let\@evenhead\@empty
 \def\@oddfoot{}%
 \let\@evenfoot\@oddfoot}
\makeatother

\usepackage{graphicx,bm}
\usepackage{lineno}
\usepackage{amsmath}
\usepackage{amssymb}
\usepackage{comment}
\usepackage{hyperref}
\modulolinenumbers[5]
\begin{document}
\begin{frontmatter}
\title{ChAKRA : The high resolution charged particle detector  array at VECC}
\author[adds1,adds2]{Samir Kundu\corref{cor}}
\ead{skundu@vecc.gov.in}
\author[adds1,adds2]{T. K. Rana}
\author[adds1,adds2]{C. Bhattacharya}
\ead{chandana@vecc.gov.in}
\cortext[cor]{Corresponding author.}
\author[adds1,adds2]{K. Banerjee \fnref{kaushik}}
\fntext[kaushik]{Present address: Department of Nuclear Physics, Research School of Physics and Engineering, Australian National University, Canberra, ACT 2601, Australia }
\author[adds1]{R. Pandey}
\author[adds1,adds2]{Santu Manna}
\author[adds1]{J. K. Meena}
\author[adds1]{A. K. Saha}
\author[adds1]{J. K. Sahoo}
\author[adds1,adds2]{P. Dhara}
\author[adds1]{A. Dey\fnref{dey}}
\fntext[dey]{Ex-research fellow. }
\author[adds3]{D. Gupta\fnref{dhruba}}
\fntext[dhruba]{Some work done during tenure at VECC}
\author[adds1,adds2]{T. K. Ghosh}
\author[adds1,adds2]{Pratap Roy}
\author[adds1,adds2]{G. Mukherjee}
\author[adds1]{R Mandal Saha}
\author[adds1]{S. Roy}
\author[adds1]{S. R. Bajirao}
\author[adds1,adds2]{A. Sen}
\author[adds1,adds2]{S. Bhattacharya\fnref{saila} }
\fntext[saila]{superannuated. }
\address[adds1]{Variable Energy Cyclotron Centre, 1/AF, Bidhan Nagar, Kolkata - 700 064}
\address[adds2]{Homi Bhabha National Institute, Training School Complex, Anushakti Nagar, Mumbai - 400 094, India}
\address[adds3]{Department of Physics, Bose Institute, 93/1 Acharya Prafulla Chandra Road, Kolkata-700009, India }

\begin{abstract}
A large 4$\pi$ array of charged particle detectors has been developed at Variable Energy Cyclotron Centre to facilitate high resolution charged particle reaction and spectroscopy studies  by  detecting event-by-event the charged reaction products emitted in heavy ion reactions at  energy  $\sim$ 10-60 MeV/A. The forward part ($\theta \sim  \pm $ $7^{0}$ - $\pm 45^{0}$) of the array  consists of 24 highly granular, high resolution charged particle telescopes, each of which is made by three layers  [single sided silicon strip($\Delta$E) + double sided silicon strip (E/$\Delta$E) + CsI(Tl)(E)]of detectors. The backward part of the array consists of 112 CsI(Tl) detectors which are capable of detecting primarily the light charged particles (Z $\le$ 2) emitted in the angular range of $\theta \sim  \pm $ $45^{0}$ - $\pm 175^{0}$. The extreme forward part of the array ($\theta \sim  \pm $ $3^{0}$ - $\pm 7^{0}$) is made up of 32 slow-fast plastic phoswich detectors that are capable of detecting  light  (Z $\le$2) and heavy charged particles (3 $\le$ Z $\lesssim$ 20) as well as handling  high count rates. The design, construction and characterization of the array has been described.
\end{abstract}

\end{frontmatter}

\section{Introduction}
The exploration of nuclear reaction dynamics is crucially dependent on the extent of information one can obtain about the reaction products; the more information one has about the reaction products, easier is the kinematic reconstruction of reaction events and deeper is the understanding of the reaction process. This is particularly important for heavy ion induced reactions at higher energies (above fermi energy), where multiplicity of reaction products is large. Event-by-event detection of maximum possible number of reaction products is necessary to enable faithful reconstruction of the events and characterise the reaction processes. Similarly, kinematic reconstruction technique also plays an important role for the study of structures of highly excited  states (e.g., unbound resonances) in nuclei. To enable such measurements, one requires a large, granular array of  detectors which is capable of detecting ideally all types of emitted charged particles with sufficient precision over the whole solid angle, 4$\pi$. It is therefore not surprising that such detector arrays (e.g., INDRA \cite{INDRA:NIMA:357:1995:418}, HIRA \cite{HIRA:NIMA:583:2007:302}, CHIMERA\cite{CHIMERA:1995}, LASSA \cite{LASSA:473:2001:302} etc.) serve as the main workhorse for reaction studies at all large accelerator centres across the world, and efforts are still continuing to build new (and/or upgrade old) detector arrays with higher granularity and better resolution (e.g. FAZIA \cite{Fazia:EPJA:50:2014:47})to probe more deeply into the nuclear reaction scenario.

The integrated experimental nuclear reaction programme based on the low ($\lesssim$ 10MeV/A) and intermediate energy ( $\sim$10-60MeV/A) heavy ions accelerated by the present K130 cyclotron and the upcoming K500 superconducting cyclotron   at Variable Energy Cyclotron Centre (VECC), Kolkata, necessitated the development of one large, granular, high resolution charged particle detector array. The detector elements are required to have large dynamic range to cope up with experiments using both the accelerators. Moreover, the design should be flexible enough to enable either a part or whole of the detector system to be used in conjunction with other types of detector systems. With due consideration of the above, a  $\textbf{Ch}$arged particle detector $\textbf{A}$rray for $\textbf{K}$inematic $\textbf{R}$econstruction and $\textbf{A}$nalysis ($\textbf{ChAKRA}$) has been developed at VECC, which is capable of detecting a wide range of emitted charged particles ($ 1\leqslant Z \lesssim 20  $)  over $\sim$ 4$\pi$ solid angle. The design details, fabrication  and characterisation of  various components of the detector array have been reported in the present paper.

\section{Design details of the ChAKRA \label{array}}
\begin{figure}
\centering
\includegraphics[scale=0.48,clip=true]{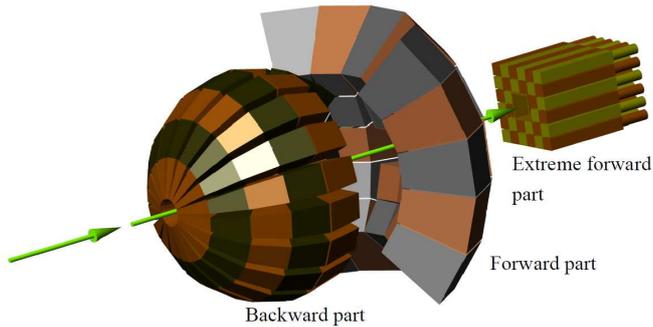}
\caption{\label{fig1:CPDA:model} Schematic view of ChAKRA. }
\end{figure}

Depending upon the reaction kinematics, yield of different types of particles will vary with angles.  Therefore, the type and granularity of the detectors should also vary with the angular range. According to that, the array has been configured in three independent blocks, all of which are  sections of  concentric spheres of different radii with the target position as the centre.  A schematic view of the charged particle detector array is shown in Fig.~\ref{fig1:CPDA:model}. The coverage of  each block is the surface area of the respective sphere spanned by two radial lines between  $ \theta_{min} \leq \theta \leq \theta_{max} $, where $ \theta $ is the angle subtended by the radial line with the beam direction (see Fig. \ref{fig1:CPDA:model}).  The most important part of ChAKRA is the forward array block, which covers the  angular range of $\theta \approx  \pm  7^{0} - \pm45^{0}$. Two other blocks are, the backward array (angular coverage : $\theta \approx  \pm 45^{0}-\pm175^{0}$), and the extreme forward array (angular coverage : $\theta \approx \pm 3^{0} - \pm 7^{0}$).  The forward array  consists of 24  charged particle telescopes, each of which is made up of 3  detector elements (Si(strip) - Si(strip) - CsI(Tl)). The backward array is  made up of  112 CsI(Tl) detectors of varying shapes and thicknesses. The extreme forward array consists of 32 plastic fast-slow phoswich detectors.
\subsection{Forward part}
\begin{figure}
\centering
\includegraphics[scale=0.5,clip=true,angle=90]{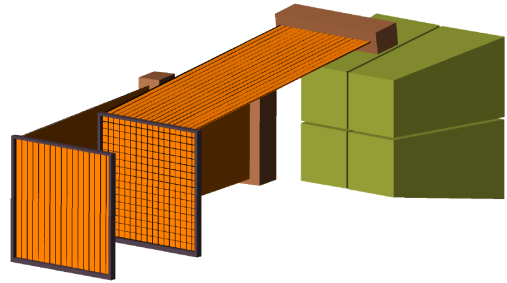}
\caption{\label{fig2:CPDA-Tel} 3D schematic diagram of charged particle telescope used in the forward array. The left most detector is SSSD of thickness $\sim$ 50 $\mu$m, middle one is DSSD of thickness $\sim$ 500 $\mu$m or $\sim$ 1000 $\mu$m and right most detectors is an assembly os 4 CsI(Tl) detectors of thickness 6 cm. }
\end{figure}

The basic requirement of the forward part of the  array is that  it should be capable of measuring both position and energy  as well as identifying all light charged particles (LCP : $ Z $= 1, 2) and intermediate mass fragments (IMF) in the range 3$ \leqslant Z \lesssim   $20 emitted in the reaction. In addition, the array should also enable isotopic identification of all LCPs and IMFs at least up to $ Z \lesssim $10. To achieve these,
single-sided Si strip detector (SSSD) of thickness $\sim$ 50 $\mu$m has been used as the outermost (facing the beam) $\Delta$E detector, and double-sided Si strip detector(DSSD)  of thickness $\sim$ 500 $\mu$m / 1 mm    has been used as E detector in all 24 telescopes.  Typical energy resolution of these detectors was $\lesssim$1 $\%$. To take care of energetic LCPs and lighter IMFs which do not stop in the Si-detectors,  an extra layer of CsI(Tl) detector of thickness 6 cm  was added behind the DSSD. The thickness of the CsI(Tl) detector was chosen to stop the highest energy light particles ($\sim$ 150MeV proton) expected in reactions at medium beam energies ($\approx$ 60MeV/A) from K500 superconducting cyclotron. Active area of each strip detector was 5$\times$5cm$^2$ with individual strip dimension of 3$\times$50 mm$^2$. SSSD has 16 strips in front (junction) side and DSSD has 16 strips on both sides in mutually orthogonal direction as shown in Fig. \ref{fig2:CPDA-Tel}. Each pixel in DSSD (size : 3$\times$3 mm$^2$) provides an angular resolution of $\sim$ 0.8$^0$ when kept at a distance of 20 cm from the target centre. Four CsI(Tl) detectors (each of front surface area 2.5$\times$2.5 cm$^2$ and thickness 6 cm) have been placed behind the thicker strip detector to complete each telescope assembly. 6  cm CsI(Tl) itself can stop proton with maximum energy $\sim$ 145 MeV. Total active area of four CsI(Tl) will cover all the particles which are not stopped by DSSD. The configuration of one telescope is shown schematically in Fig. \ref{fig2:CPDA-Tel}.
\subsubsection{Forward part Si-strip detectors}
\begin{figure}
\centering
\includegraphics[scale=0.42,clip=true]{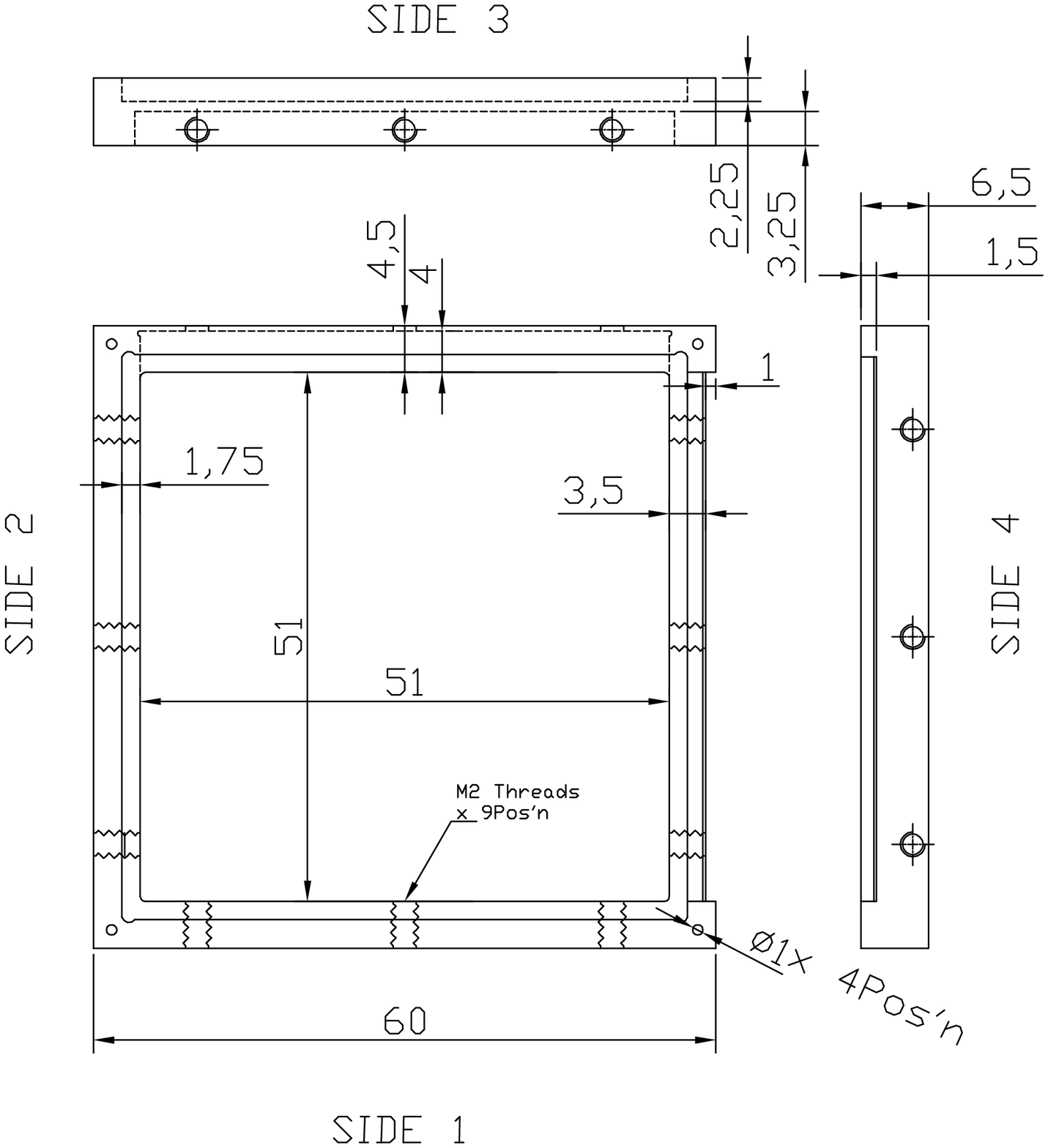}
\caption{\label{fig5-10:Strip-frame} Design of the DSSD frame. All dimensions are in mm. }
\end{figure}
\begin{figure}
\centering
\includegraphics[scale=0.4,clip=true]{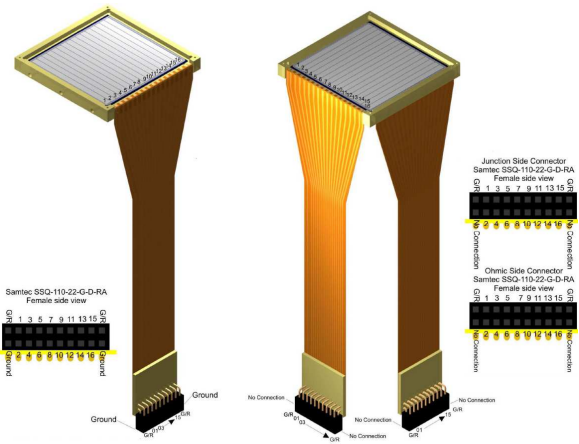}
\caption{\label{fig5-11:Strip-detector} Strip detectors with kaptons.}
\end{figure}
The silicon strip detectors (SSSD and DSSD) used in the present array are ion-implanted, passivated  devices obtained from M/s. Micronsemiconductors Ltd., UK \cite{micron}. Total dead layer in the detector is typically  $\sim$ 0.6 $\mu$m including the implantation depth of about $\sim$ 0.40 $\mu$m. Typical energy resolutions of the detectors (individual strip) are, $\sim$ 40 keV for the (50 $\pm$5)$\mu$m SSSD and  $\sim$ 40 keV ($\sim$ 25 keV) for  500 $\mu$m (1000 $\mu$m) DSSD, measured using 5.486 MeV  $^{241}$Am-$\alpha$ source. The frames of the detectors were custom designed (see Fig.~\ref{fig5-10:Strip-frame}) to fit into the present array structure to minimise the dead area (frame size) surrounding each silicon wafer. The frame is made from glass epoxy with total outer dimension 60 mm $\times$ 60 mm [Fig.~\ref{fig5-10:Strip-frame}].  A slot of width  1.75 mm and depth 2.25 mm in the inner (top) sides of the frame [Fig.~\ref{fig5-10:Strip-frame}] helps to glue the silicon wafer on the frame as well as protect the wafer from other detectors when used in telescopic mode. The outer ridge has four through holes, one in each corner, which can be used to align the frame by dowel pins. Signals from the strips are taken out by kapton cables with 18 soft gold pads of width 1.6 mm of each, the other end of which is connected to a flat ribbon connector(FRC) (SAMTEC SSQ-22-G-D-RA)  (see Fig.~\ref{fig5-11:Strip-detector}). 16 pads are used for 16 signals connections and 2 for guard rings. The slots in side 4 ensure proper spacings for wire bonding of the front side strip with the kapton and its  ($\Delta$E detector) passage through the spacing between E detector's frame and housing wall.  The slot in side 3 allows  the kapton to be connected to backside strip (in DSSD). Both SSSD  and DSSD frames are same in all respect except that there is no kapton in side 3 (Fig.~\ref{fig5-11:Strip-detector}) for SSSD.
\subsubsection{Forward part CsI(Tl) detector}
\begin{figure}
\centering
\includegraphics[scale=0.53,clip=true]{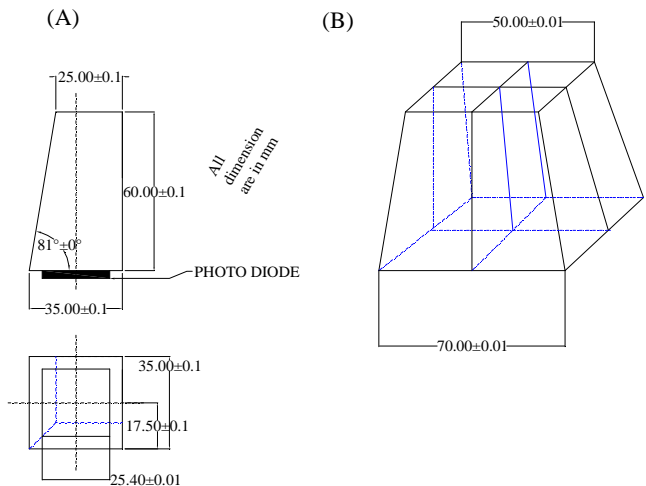}
\caption{\label{fig5-14:CsI-design} (A) Design of one CsI(Tl) detector and (B) assembly of four CsI crystal as used in the telescope.}
\end{figure}

\begin{figure}
\centering
\includegraphics[scale=0.7,clip=true]{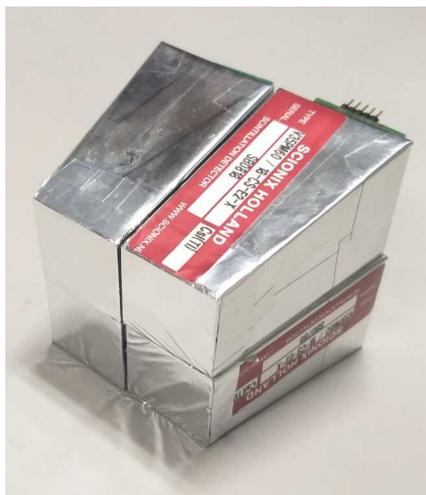}
\caption{\label{fig5-14c:CsI-stack} Assembly of four CsI(Tl) detectors for forward array telescope.}
\end{figure}

Thallium activated Cesium Iodide [CsI(Tl)] detectors constituted the third element of the forward array telescopes. All the CsI(Tl) detectors have been developed by M/s. Scionix Holland BV \cite{Scionix} as a complete assembly of CsI(Tl) crystal and photo diode (Hamamatsu S3204-08) with a charge sensitive preamplifier \cite{Scionix}. Custom made design of the detector has been done in association with M/s. Scionix Holland BV. The forward array has been designed in  such a way  that the front face of the first strip detector formed a part of the surface of a sphere of radius $\sim$ 20 cm. Accordingly, the shape of the CsI(Tl) crystal has been decided  such that the trajectories of the emitted particles were fully contained within the crystal. The design of the single detector, each of thickness 6 cm, is shown in Fig.~\ref{fig5-14:CsI-design}(A); the front and back faces are square shapes of dimensions 2.5 cm $\times$ 2.5 cm, and 3.5 cm $\times$ 3.5 cm, respectively. The thickness of CsI(Tl) is sufficient to stop a proton with  energy $\sim$ 145 MeV. We have used an assembly of four such detectors in the third layer of the telescope; the segmentation provides    better energy resolution and identification of highest energy lighter particles. The assembly of four CsI(Tl) forms a truncated pyramid of base area 7 cm $\times$7 cm and its front face is having same area as the active area of the strip detectors as shown in Fig. ~\ref{fig5-14:CsI-design}(B). Each crystal has been wrapped in a special reflecting material covered with aluminized mylar of thickness $\sim$ 50 $\mu$m on all sides except the front face. This will ensure better light collection and stop light leakage to adjacent detector. The front face has been covered with ultrathin ($\sim$ 1-2$\mu$m) aluminized mylar sheet to allow the minimum dead layer for the incoming charged particles. Typical energy loss of 100 MeV $\alpha$ in $\sim$ 1-2$\mu$m aluminized mylar is $\sim$ 70-140 KeV \cite{LISE}.  The crystals were coupled with  photodiodes of active area 18 mm $\times$ 18 mm (Hamamatsu S3204-08). A charge sensitive preamplifier \cite{Scionix} with gain of $\sim$4V/pC was directly coupled to the photodiode.
A stack of four such detector are shown in Fig.~\ref{fig5-14c:CsI-stack}
\subsubsection{Forward part telescope housing}
\begin{figure}
\centering
\includegraphics[scale=0.49]{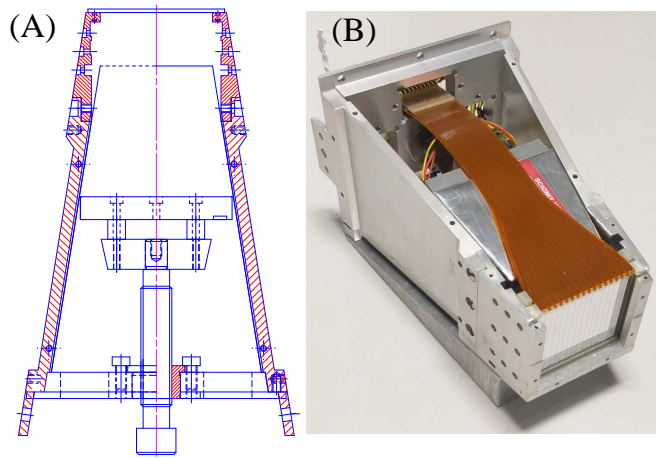}
\caption{\label{fig5-17:Housing} (A) Design of a telescope, and (B) photograph of the telescope with strip detectors and CsI detectors.}
\end{figure}

The design of the telescope housing and its actual photograph after fabrication are shown in Fig.~\ref{fig5-17:Housing}. A special hardened  Aluminium alloy (6061-T6) has been used for fabrication of these housings. This alloy is easily machinable with high precision. The housings were designed in such a way that the telescope will contribute minimum dead area. The housing has two parts. In the front part, two strip detectors are kept in a fixed condition in two parallel slots. A provision is there to fix the strip detectors rigidly with the housing using screw. To keep the four CsI(Tl) detectors behind the strip detector (E), inner volume of the housing has been given a shape according to the overall shape of four  CsI detector assembly, i.e., truncated pyramid shaped. In the inner sides of the housing there are slots of depth 0.8 mm and width 44 mm to take out the kapton of the strip detectors. To hold four CsI(Tl) detectors in proper position, there is an arrangement by a long threaded rod as shown by Fig. ~\ref{fig5-17:Housing}. By rotating the rod slowly, the detectors can be placed in proper position. The outer shape of the housings is such that when all the housings will be kept in the form of array, it will form a part sphere of radius 20 cm. To attach  all the housing in the array, a deep slot was kept outside of each four walls of the housings. A complete assembly of the housing is shown in Fig. ~\ref{fig5-17:Housing} (right) with top open.
\subsubsection{Forward part support structure}
\begin{figure}
\centering
\includegraphics[scale=0.45]{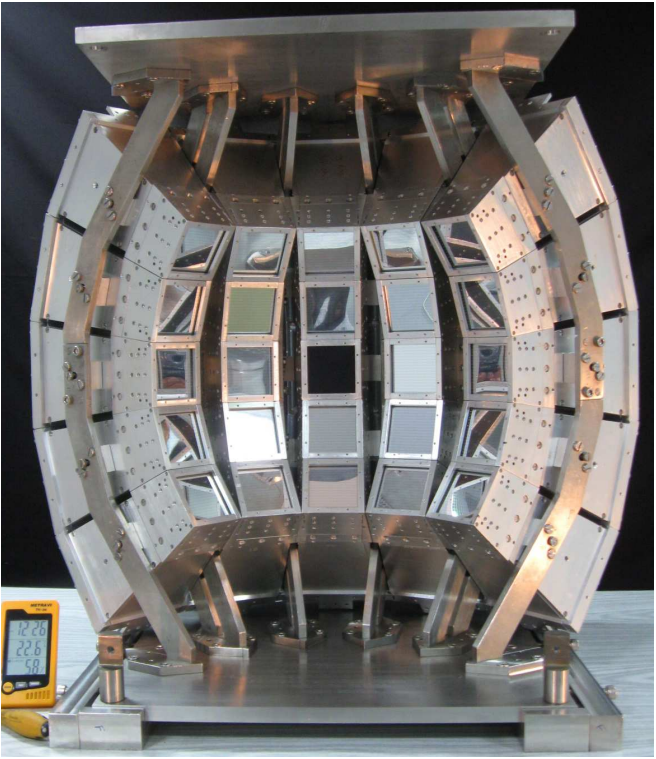}
\caption{\label{fig5-18:Array} Forward array of charged particle telescopes. }
\end{figure}

To mount the telescopes in a form of array, a support structure was developed. The telescopes have been arranged in 5 columns with five telescopes in each except in the  middle column, where the central telescope has been removed for the passage of beam. Shapes of the columns  are such that, all the front faces will touch the surface of an imaginary sphere of radius 20 cm. In the same way, all the column have been placed so that front faces of all the telescopes together formed a part of the spherical surface of 20 cm radius spanning the angular range of $7^{0} \le \theta \le 45^{0}$.   The whole array can be  moved both vertically and horizontally to align it with the beam line axis. The complete assembly of 24 telescopes has been shown in Fig.~\ref{fig5-18:Array}.
\subsection{Backward array\label{backward:sec}}
Since very few IMFs are emitted in the backward angular range (beyond the zone covered by the forward array), it has been decided to detect mainly the LCPs in the angular range $\theta$  = 45$^0$-175$^o$ using in backward part of ChAKRA. So, the backward part has been made up of  112   CsI(Tl) detectors of various sizes as described below. Pulse shape discrimination (PSD) technique \cite{GUINET:NIMA:278:1989:614} has been used to identify the LCPs.
 \begin{figure}
 \centering
    \includegraphics[scale=0.4,clip=true]{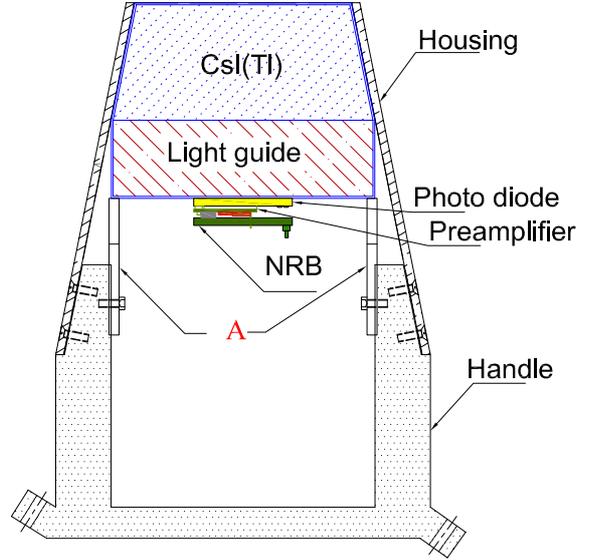}
   \caption{\label{back-housing}{Schematic of Backward array CsI(Tl) detector with housing}}
\end{figure}
\begin{figure}
\centering
    \includegraphics[scale=0.4,clip=true]{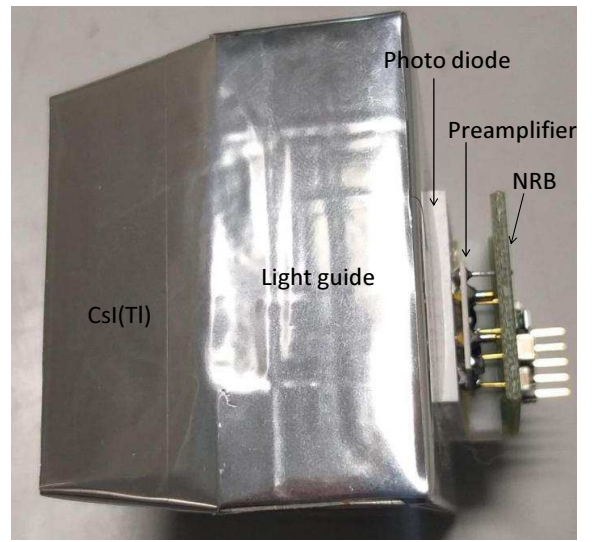}
   \caption{\label{CsI-back}{Typical backward array CsI(Tl) detector with preamplifier and noise reduction board (NRB).}}
\end{figure}
\subsubsection{Backward part detectors and support structure}
The shapes of the detectors of this array are such that the front faces of the detectors form the part of a spherical surface of radius ~15 cm in the angular range $\theta$  = 45$^0$-175$^o$. The shapes and sizes of the detectors were optimised for minimum dead area and multihit probability. Each crystal, trapezoidal in shape, has been wrapped in a special reflecting material covered with $\sim$ 50$\mu$m thick aluminized mylar foil on all sides to prevent cross talk of scintillation light between the adjacent detectors. The front face has been covered with ultrathin ($\sim$ 1.5$\mu$m) aluminized mylar. This allow the minimum dead layer for the incoming charged particles. Each detector was coupled, either directly or through light guide depending upon the thickness. It helps good light collection in the case of thinner detector (ring-3 to ring-6). All the crystals are coupled  with photodiodes and vacuum compatible charged sensitive preamplifiers, forming an integrated detector assembly. The complete detector assembly, as per the design specifications, was fabricated by M/s. Scionix Holland Bv \cite{Scionix}.  The size of the photodiode varied with the size of the detector. The design of a typical detector with housing and the photograph of a complete detector are shown in Figs.~\ref{back-housing} and \ref{CsI-back}.  In all, nine types of detectors of different shapes and thickness have been used in the array. The type of the detectors are named according to the ring number on which they are kept.

\begin{figure}
\centering
    \includegraphics[scale=0.45,clip=true]{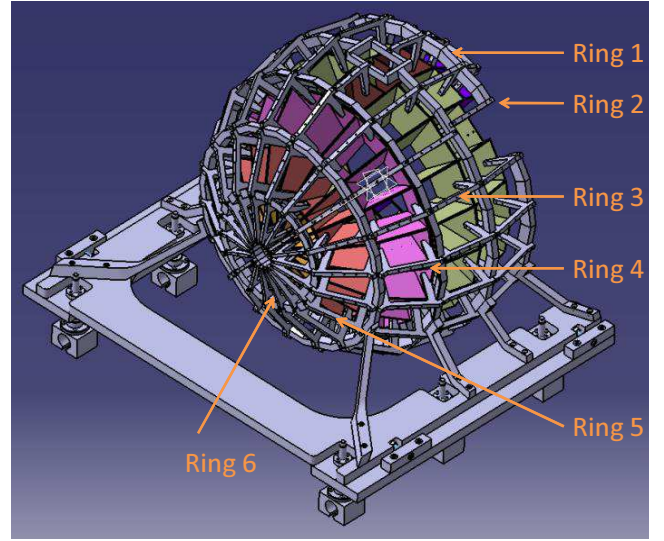}
   \caption{\label{cad-back}{3D CAD model of support structure and housings of backward part of ChAKRA}}
\end{figure}

The design of the array support structure has been illustrated  in Fig. \ref{cad-back}. The backward array consists of six azimuthally symmetric rings. The details of the detectors and their placement in different rings is elaborated in Table \ref{table1}. The ring-1 is partial and houses six detectors. Provision for inserting the  target ladder has been kept between ring-2 and ring-3. All the other rings have eighteen detectors of a single type of detector in each ring (see Table~\ref{table1}). The design of the array(Fig. \ref{cad-back}) is such that each of the housing (Fig. \ref{back-housing}) can be removed individually without disturbing the other detector and housings. At the end of the housing handle, two screws are there to attach the housing with the support structure. To fix the detector inside the housing two plates (marked with`A' in Fig. \ref{back-housing}) are there which can be fixed on the handle with screw.
\begin{table}
\centering
\caption{\label{table1} Details of the CsI(Tl) detectors at different rings. Rn $\rightarrow$  ring number, Th $\rightarrow$  detector thickness (in cm), Type $\rightarrow$ Type of  detector according to shape, Q $\rightarrow$ Quantity of each type of detectors in any ring. The  A $\rightarrow$ Area (cm$^2$) of the front face of the detector,  PD $\rightarrow$ Active area (cm$^2$) of photo diode and   $\theta_{1}$,  $\theta_{2}$ defines the angular width of each ring. Each detector covers an azimuthal width $\Delta\phi$ of 20$^0$.}
\begin{tabular}{cccccccccc}
\\
\hline
 Rn         & Th& Type   & Q              &  A     &PD &$\theta_{1}$ & $\theta_{2}$       \\
\hline
\\
1           & 4  & CsI-1  & 6             &12.12 & 3.24 & 54.3   & 67.5 \\
2           & 4  & CsI-2  & 24            &12.55 & 3.24 &67.5    & 90\\
            &    & CsI-2A & 2             &12.23 & 3.24 &67.5    & 73\\
            &    & CsI-2B & 8             &8.82  & 3.24 &75.1    &90.0\\
3           & 3  & CsI-3  & 16            &25.57 & 3.24 &90.0    &112.5\\
            &    & CsI-3A & 2             &12.23 & 3.24 & 100.4  & 112.5\\
4           & 2  & CsI-4  & 18            &21.23 & 3.24 &112.5   & 135.0\\
5           & 2  & CsI-3  & 18            &13.08 & 2.00 &135.0   &157.5\\
6           & 2  & CsI-4  & 18            &2.87  & 1.00 & 157.5  &170.4\\
\hline
\\
\end{tabular}
\end{table}
 \begin{figure}
 \centering
    \includegraphics[scale=0.45,clip=true]{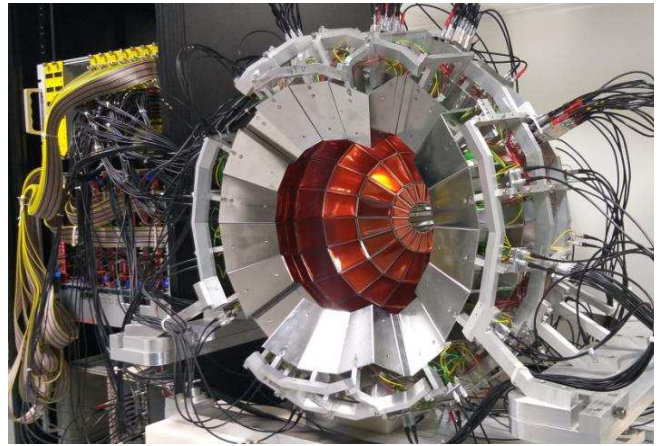}
   \caption{\label{cpda-back}{Backward part of ChAKRA with 112 CsI(Tl) detectors and its electronics.}}
\end{figure}
\subsection{Extreme forward array}
\begin{figure}
\centering
    \includegraphics[scale=0.5,clip=true]{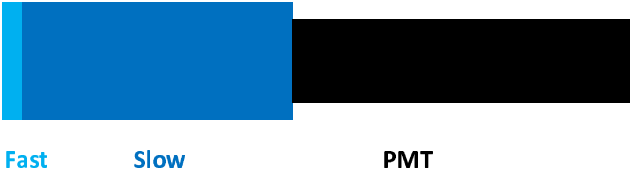}
   \caption{\label{plastic-schematic}{Schematic diagram of plastic phoswich detector with PMT. }}
\end{figure}

\begin{figure}
 \centering
    \includegraphics[scale=0.25,clip=true]{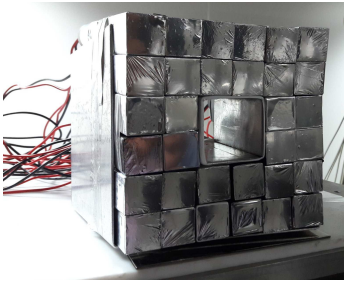}
   \caption{\label{extreme-cpda-photo}{The array of 32 plastic phoswich detectors with a square shaped pipe with area of four detectors.}}
\end{figure}

In the extreme forward angle (in the neighbourhood of grazing angle), detectors should be able to handle high counting rate and they should also be rugged against radiation damage due to high rate of bombardment. Plastic Scintillator is the quite suitable to handle such situation quite efficiently.
So, the extreme forward array has been designed with 32 plastic phoswich detectors. Each phoswich detector was made up of 100 $\mu$m (2ns) thin organic fast plastic scintillator (BC 408) ($ \Delta $E-detector) and 10 cm slow (280ns) thick plastic scintillator (BC 444) (E-detector) as shown in Fig.\ref{plastic-schematic} schematically. Active area of each detector is 20$\times$20 mm$^2$ and it is  coupled to a photomultiplier (Hamamatsu PMT model R6511) of 19 mm diameter. The array of these detectors is placed at 40 cm from the target which will cover an angular range of $\theta$ $\sim$  3$^0$ to 7$^0$. In the middle of the array, a square shaped pipe with area of four detectors, has been provided for transporting the beam to the beam dump.
All plastic phoswich detectors of the array have been fabricated at VECC. Thick part of  plastic  used in  the detector has been developed by M/s. Saint-Gobain Crystals\cite{saint-gobain} in proper shape. Thin plastic (BC 408) detectors of area 20$\times$20 mm$^2$ have been prepared from large sheet of area 10$\times$10 cm$^2$. Both this parts have been pasted with the optical cement, SC-500 \cite{Scionix} and then painted with a white reflecting paint, SC 510 \cite{Scionix}, to prevent scintillation light loss. The thin end of the detector has been covered with 1.5 $\mu$m aluminized mylar foil and a photo multiplier tube has been optically coupled at the other end to make the phoswich detector. All sides of the plastic (except front) as well as the PMT joint were wrapped with a 50 $\mu$m thick  aluminized mylar foil to avoid scintillation light leak. The extreme forward array of plastic phoswitch detectors is shown in Fig.\ref{extreme-cpda-photo}.
\subsection{Readout electronics and data acquisition system}
The readout of the array comprised of standard NIM/CAMAC front-end signal processing electronics followed by the VME Data Acquisition System (VMEDAQ). We have used 16 channel preamplifiers (MPR-16) and shaping amplifier cum constant fraction discriminator modules  (MSCF-16)from M/s. mesytec GmbH \& Co. KG \cite{mesytec} for the detectors used in the forward part of ChAKRA.The outputs of the shaping amplifiers are connected to 32 channel peak sensing VME ADC (Model CAEN V785). For backward part of ChAKRA, we used fast gate - slow gate charge integration method to identify the particle and measure of its energy. The output of the preamplifier coupled with CsI(Tl)-photodiode was fed to the shaping amplifier cum leading edge discriminator module  (MSCF-16)\cite{mesytec} specially designed by M/s. mesytec GmbH \& Co. KG for the present purpose. Charge integration was done using 32 channel VME QDC (Model CAEN V792). Double gate charge integration  was also used for particle identification in the plastic phoshwich detectors of the forward part of ChAKRA. Here, the output of the PMT is directly amplified by a fast amplifier (CAEN N412) having two outputs in each channel . One output of each channel has been divided into two parts  using a splitter circuit. The time trigger was generated using 16 channel CAMAC leading edge  discriminator (LED) (CAEN C894). OR of two such LED are again ORed using 4 channel logic fan-in fan-out module having 4 input 4 output in each channel. Finally it goes to VME QDC (CAEN V792) through gate and delay generator (Ortec 8020)for fast gate- slow gate charge integration. Each ADC (QDC) module  digitized the peak values (integrate charge) of all 32 channels simultaneously within 5.6 $\mu$s with 12 bit resolution and stores the data into FIFO. The VMEDAQ software reads the data in VME chain block transfer (CBLT) mode from multiple crates and builds the global events. The software parallelizes the operations like reading, event construction, storing and processing  of the data using multiple threads in a producer-consumer model. The internal FIFO of the ADCs/QDCs and the multi-threaded parallel operations on a multi-core workstation are essential features to improve the dead time of the system. The current version of the data acquisition operates in common dead time mode. The common GATE signal is used for all the crates. A custom made sychronizer module is used to block the GATE signal, until all the ADC/QDC busy signals are withdrawn. This ensures the synchronization of the data acquisition by all the ADCs. The event correlation is achieved through hardware synchronization. Each event is marked with a GATE count value. This GATE count is used to correlate the events from different crates. The future upgradation of the VMEDAQ will also support the timestamping mechanism to correlate the events.
\section{Characterisation of ChAKRA \label{charac}}
Detailed offline and in-beam tests have been carried out to characterise various elements of ChAKRA. Initially, offline tests using radioactive sources have been done to check the basic parameters, like, energy resolution, thickness uniformity  as well as uniformity of spatial response from the detectors for acceptability. In the next step, they have been tested in real in-beam experimental situation to test the performance.
\subsection{Forward array}
\begin{figure}
\centering
\includegraphics[scale=0.8,clip=true]{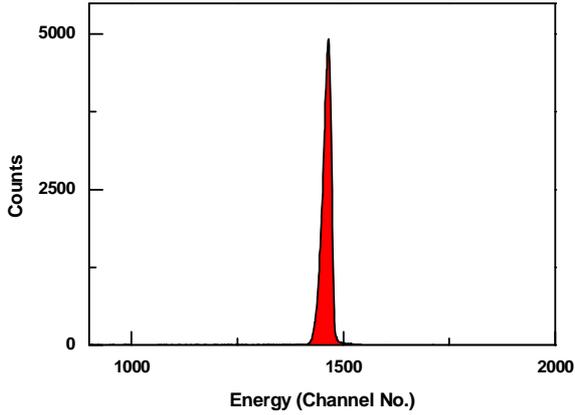}
\caption{\label{fig5-12:Strip-E-resolution} The energy spectrum of $\alpha$-particles emitted from  $^{241}$Am  source obtained by an individual strip. }
\end{figure}
\subsubsection{Characterization of silicon strip detectors \label{uniformity1}}
The strip detectors have been tested in offline (with $\alpha$-sources) as well as in beam. Offline characterisation  of each individual strip has been done using a collimated pin-type $^{241}$Am  5.486 MeV $\alpha$-sources specially prepared in-house for this purpose. The pin source (diameter 1 mm , collimated with a tube of length 5 mm) was kept very close ($\sim$ 2 mm) to the detector to irradiate the detector in a very small area (circle of diameter 1.4 mm) in a particular position. The best energy resolutions of individual strips  for 50 $\mu$m, 500 $\mu$m and 1 mm detectors have been found to be  45, 40 and 25 KeV, respectively. Typical energy spectrum of $\alpha$-source in one of the strips is shown in Fig.~\ref{fig5-12:Strip-E-resolution}.

\begin{figure}
\centering
\includegraphics[scale=0.4,clip=true]{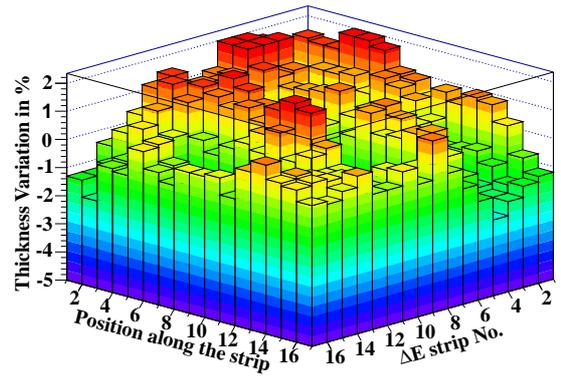}
\caption{\label{fig2-7:nonunif} Non-uniformity in thickness along the strip in 16 different positions for all strips of a 50 $\mu$m SSSD.}
\end{figure}

Thickness variation along a strip of the detector, particularly in 50 $\mu$m thin SSSD's, which are very prone to thickness non-uniformity,  results in poor isotopic identification. So, it is very crucial to measure the thickness non-uniformity, particularly for thin detectors. Typical thickness non-uniformity of a thin (50 $\mu$m SSSD) detector, estimated in an in-beam experiment, has been shown in Fig. ~\ref{fig2-7:nonunif}. We have used $\alpha$ emitted in reaction 145 MeV $^{20}$Ne + $^{12}$C using same method described in \cite{LASSA:473:2001:302}. It has been found that the variation in thickness along a strip is  $\lesssim$ 3 $\%$ of the average thickness.
\subsubsection{Characterisation  of forward part CsI(Tl) detector \label{csitest}}
\begin{figure}
\centering
\includegraphics[scale=0.6]{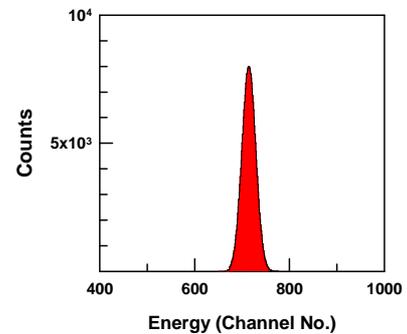}
\caption{\label{fig5-16:CsI-EResolution} Spectrum of $\alpha$-particle emitted from $^{241}$Am measured using CsI(Tl) detectors.}
\end{figure}

The detailed characterisation of all CsI(Tl) detectors have been done using $^{241}$Am $\alpha$-source. Typical energy spectrum of the detectors has been displayed in Fig.~\ref{fig5-16:CsI-EResolution} and the energy resolution was found to be $ \sim $ $5\%$. Depending upon the uniformity of the thallium doping and the geometry of the detector, there may be a  non-uniformity in the light output from different regions of a single CsI(Tl), which may cause variation in isotopic resolution between different pixels of DSSD and CsI(Tl) of the forward array telescope.

\begin{figure}
\centering
\includegraphics[scale=0.45,clip=true]{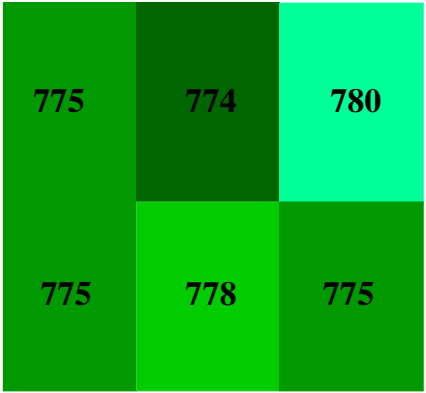}
\caption{\label{fig5-15:CsI-nonuniformity} Non-uniformity in the light output with respect to the mean peak position (776). Numbers are the peak position of the $\alpha$-spectrum in terms of ADC channel number. }
\end{figure}

To measure the non-uniformity, a pin-type $\alpha$-source has been put at different positions very close (2 mm) to the front face of the detector. Fig.~\ref{fig5-15:CsI-nonuniformity} represents  the front face (area 25 mm $\times$ 25 mm) of a CsI(Tl) detectors. Numbers are the peak positions of the $\alpha$-spectra in terms of ADC channel number at the centre of each block where the source has been kept during the test. With respect to the mean peak position, the variation was found to be  $\sim \pm$0.49$\%$.
\subsubsection{Particle identification in forward part telescope}
\begin{figure}
\centering
\includegraphics[scale=0.4]{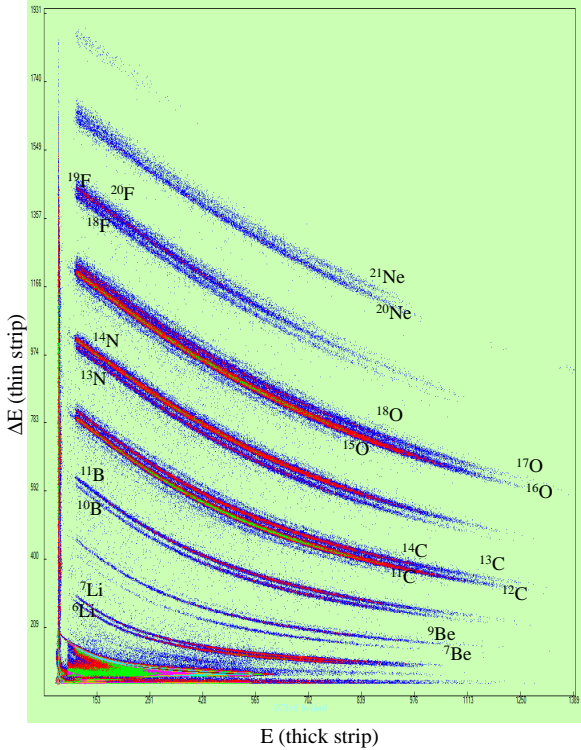}
\caption{\label{fig5-16b:2D-Spec} 2D Spectrum of intermediate mass fragments and their isotopes emitted in the reaction 193MeV $^{20}$Ne on $^{9}$Be obtained by a telescope.}
\end{figure}
\begin{figure}
\centering
\includegraphics[scale=0.5]{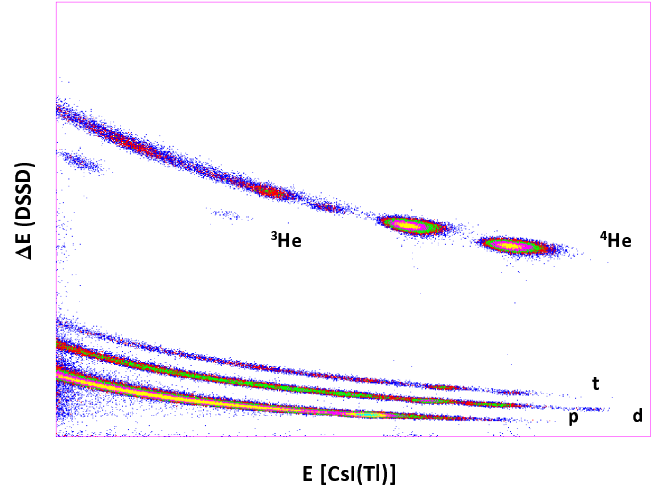}
\caption{\label{fig5-16a:2D-Spec} 2D Spectrum light charged particles emitted in the reaction 60MeV $\alpha$ on $^{12}$C obtained by a telescope.}
\end{figure}

After the fabrication of the forward telescopes, they  have been tested in beam to check their performances, the particle identification in particular. In one test experiment described here, 193 MeV $^{20}$Ne beam was incident on $^9$Be target and the fragments were detected \cite{DAE:63:2018:678}. The telescope has been kept at $\sim $ 20 cm. The angular resolution is $\sim$ $\pm$ 0.8$^0$. The $ \Delta $E-E  spectrum of the emitted fragments   obtained from  first two layers [Si(SSSD)-Si(DSSD)] of the telescope is shown in Fig. \ref{fig5-16b:2D-Spec}. Clear isotopic separation of the fragments up to atomic number Z = 10 has been observed. Measurement and identification of LCPs, which punch through the two Si detector layers and deposit energy in CsI(Tl) detector at the back end of the telescope, require the usage of  Si-CsI(Tl) component of the telescope. A typical LCP study in the experiment of 60 MeV $\alpha$ on $^{12}$C target is presented here, where  the $ \Delta $E-E spectrum obtained by last two detectors of the 3-element telescope, i.e., DSSD and CsI(Tl) detector is shown in Fig. ~\ref{fig5-16a:2D-Spec} \cite{Tapan:PRC:78:2008:027602}. It is seen that all isotopes of Z =1 and 2 are clearly identified. These measurements clearly establish the expected level of performance of the array. The telescope is able to isotopically separate all fragments at least up to Z=10, which was the bench mark of performance.
In actual physics experiment, one  encounters multiple hit events. Therefore, kinematic reconstruction of the events (identification of all particles, and their pixel positions and energies) is essential to extract the physics information. Here, a pixel corresponds the overlapping area between the orthogonal  front and back side strips of the DSSD detector (3$ \times $3~mm$ ^2 $ in the present case).  A data reduction algorithm in ROOT framework \cite{ROOT} has been developed to identify the particles and reconstruct the events. The hit positions of the particles are identified by matching the energies deposited in the front and back strips, using a minimization routine specially developed for this purpose.
\subsection{Characterization of the backward part detectors}
\begin{figure}
 \centering
    \includegraphics[scale=0.9,clip=true]{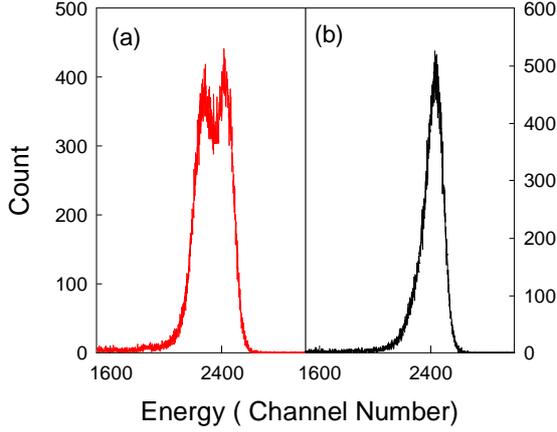}
   \caption{\label{E_res_back}{Energy spectra $^{241}$Am $\alpha$ source using CsI(Tl) detectors. The energy spectra of panel (a) was obtained from a detector having defective entrance window foil and panel (b) after replacement of the foil.}}
\end{figure}

 Non-uniformity in doping of CsI(Tl) crystals causes non-uniformity in scintillation light output which affects the energy resolution. Non-uniformity has been measured in a similar way as mentioned Sec. \ref{csitest}, which has been found to be within 1$\%$. Energy resolution of detectors also depends on the variation of the thickness and micro-leak in the entrance window.  The energy resolution of each detectors has  been measured by using a $\alpha$ source of $^{241}$Am having energy 5.486 MeV. In case of some of the detectors we found two peaks in the  energy spectra from $^{241}$Am $\alpha$ source as shown in Fig. \ref{E_res_back}(a). After replacement of the mylar window, one peak disappeared as in Fig. \ref{E_res_back}(b). The extra peak was coming  due to variation of thickness in the entrance mylar window. It was observed that the energy resolution of the detectors, depending on the volume of the crystal, varied between $\sim$ 4-6 $\%$.
In the backward array, light particles are identified by using the pulse shape discrimination (PSD) property of the CsI(Tl) detector. The light output
 \begin{figure}
 \centering
    \includegraphics[scale=0.52,clip=true]{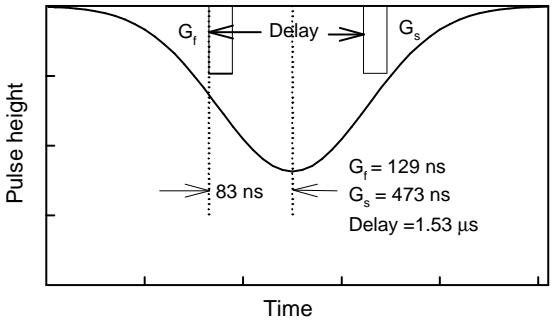}
   \caption{\label{psd_gauss}{Schematic diagram of fast gate and slow gate.}}
\end{figure}
of CsI(Tl) detectors have two components which can be approximately written as
\begin{equation}
L(t) = L_{s}exp(-t/\tau_s) + L_{f}exp(-t/\tau_f)
\end{equation}
where $L(t)$ is the light output at time \textit{t}.  ($L_s$, L$_f$) and ($\tau_s$, $\tau_f$) are amplitude and  decay constant of scintillation light output of  fast and slow components, respectively. $\tau_f$ depends on the stopping power \textit{(dE/dx)} of the particle in the CsI(Tl):   $\tau_f \propto  \textit{dE/dx}$. This property is used to identify the particle using  pulse shape discrimination (PSD) technique. Particle identification property of these detectors using PSD technique has been studied in-beam experiment using 50 MeV $\alpha$ beam from K130 cyclotron,VECC with $^{27}$Al target. Here, charge integrations of the detector pulse were done over two gates, a fast gate ($G_f$) and a slow gate ($G_s$), as shown schematically in Fig \ref{psd_gauss}. The spectrum obtained is shown in Fig. \ref{psd}. It is seen that \textit{p}, \textit{d}, \textit{t} and  $\alpha$ are clearly separated \cite{DAE:57:2012:864}.
 \begin{figure}
    \includegraphics[scale=0.45,clip=true]{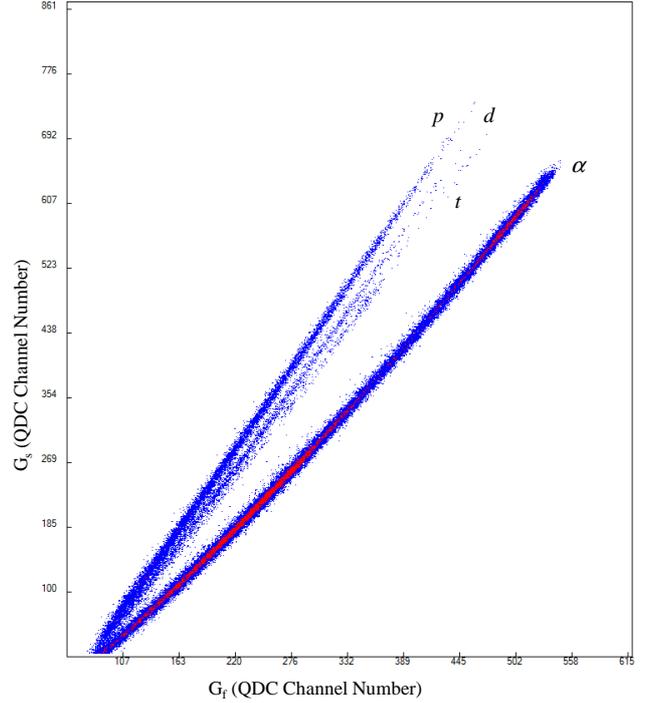}
   \caption{\label{psd}{Fast gate - slow gate spectrum for gates and delay widths as in Fig. \ref{psd_gauss}}}
\end{figure}
 At low energy, the light output (L) of a CsI(Tl) detector depends non-linearly on   the deposited energy (E). Therefore, unlike Si detectors, the energy calibration of CsI(Tl) is not quite straightforward as it depends on particle type. Therefore, the energy calibration procedure for these detectors was tested using three types of particles, $ ^7 $Li, $ \alpha $ and protons.  Elastically scattered $ ^7$Li from a gold target as well as  proton recoil peaks are used for this purpose. In addition, $ \alpha $ and protons of known energies produced in the  $^{12}$C($^{12}$C, $^{4}$He)$^{20}$Ne$^*$ and $^{12}$C($^{12}$C, $p$)$^{23}$Na$^*$ reactions at beam energy 40 and 25 MeV feeding the excited states of Ne$^*$ and $^{23}$Na$^*$ were used \cite{DAE:51:2006:618}. The light outputs were fitted separately as well as globally using two algorithms \cite{DAE:51:2006:618}.
\begin{equation}\label{eqn:calib1}
E(L) = aL + b\ln(1+cL) : individual \\
\end{equation}
and,
\begin{equation}\label{eqn:calib2}
E(L, Z,A)= a AZ^2L + (b+CAZ^2) L^{(1-dZ\sqrt{A})} : global \\
\end{equation}
Best fit parameter sets for both the algorithms are given in table.\ref{table2}
\begin{table}
\caption{\label{table2} The fitted value of the parameters used in Eqn. \ref{eqn:calib1} and \ref{eqn:calib2}}
\begin{tabular}{ccccc}
\\
\hline
 Z                  & a & b   & c      &  d       \\
\hline
p                   & 0.0916  & 0.2412 & 0.0258            & -  \\
$\alpha$            & 0.0941  & 0.7381 & 5.2631            & - \\
$^7$Li              & 0.0994  & 4.9171 & 0.0773            & - \\
p, $\alpha$, $^7$Li & 0.0008  & 0.1040 & 0.0110            &0.0455\\

\hline
\\
\end{tabular}
\end{table}
\subsection{Characterization of plastic phoswich detectors of extreme forward part}
Decay constants of the light output of these two plastics of the phoswich detector are different; for BC 408 (thin),  the decay constant is 2.1 ns \cite{saint-gobain} and the same for BC 444 (thick) is 285 ns \cite{saint-gobain}. This thin-thick combination with fast-slow decay times acts as a $\Delta E-E$ combination. The charged particle incident on the detector produces light having fast and slow components which is collected by single photo multiplier tube (PMT). Two integration of charge, one fast gate and another slow gate,  in proper time will give the light outputs proportional to   energy deposited in thin and thick plastics separately and two dimensional plot of the two outputs provide information about the type and energy of the particle.

\begin{figure}
 \centering
    \includegraphics[scale=0.5,clip=true]{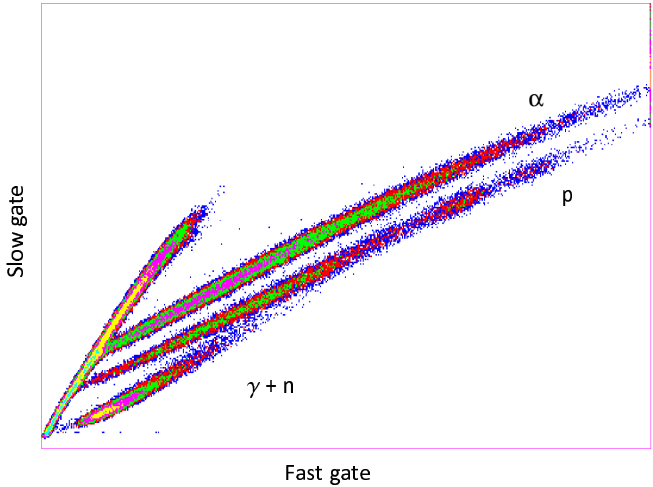}
   \caption{\label{plastic-spectrum}{Fast gate - slow gate energy spectra obtained by plastic phoswich detector}}
\end{figure}

The performance of these detectors was tested in-beam using the reaction 145 MeV $^{20}$Ne + $^{12}$C. The PMT output signal was fed into fast amplifier and amplified output  was divided into two parts. These two parts  were then charge integrated  using the two gates, each of width 60 ns and sepated by 240 ns. A two dimensional plot of the two (Fig. \ref{plastic-spectrum}) demonstrated the particle identification property of these detectors.
\subsection{Experiments using ChAKRA}
Experiment using full ChAKRA is planned with the beams from  upcoming K500 superconducting cyclotron at VECC. For that purpose, a large, segmented, horizontal axis, reaction chamber (SHARC)\cite{samir:JoP:390-2012-012075} has already been developed to house the ChAKRA.  In the meanwhile, different parts of the array, the segments of the forward array in particular, have been and are being extensively used in wide varieties of nuclear reaction studies at VECC and other accelerator laboratories across India (e.g., refs.\cite{Tapan:PRC:78:2008:027602},\cite{Tapan:88:2013:021601}, \cite{Vishal:PRC:91:2015:54611,Vishal:93:2016:044601}, \cite{Santu:PRC:94:2016:051601},\cite{ratnesh:95:2017:064603} ).
\section{Summary}
ChAKRA, the 4$ \pi $ array of charged particle detectors have been developed at VECC for medium energy ($\lesssim$ 60MeV/A) reaction studies. However, part of ChAKRA can be used for the experiments at  low ($\sim$ 10MeV/A) energy also. The array consisted of three independent blocks and each block is made up of different type of detector systems, depending on the nature of reaction products expected within the coverage of the particular block. Most important part of the ChAKRA  is the high resolution forward array (angular coverage : $\theta \sim  \pm $ $7^{0}$ - $45^{0}$) which consists of 24 charged particle telescope each of which is made by three layers of detectors [single side silicon strip($\Delta$E) + double side silicon strip (E/$\Delta$E) + CsI(Tl)(E)]. This part is the main workhorse of the array which can identify isotopes of light charged particle as well as heavy charged particles at least up to Z = 10. Backward part of the array consists of 112 CsI(Tl) detectors. It covers an angular range of  $\theta \sim  \pm $ $45^{0}$ - $175^{0}$. Pulse shape discrimination method is used for these detectors to detect light charged particles with their isotopic separation. Extreme forward part of the array consists of 32 plastic phoswich detectors, covering the most forward angular range of  $\theta \sim  \pm $ $3^{0}$ - $7^{0}$. These fast detectors are capable to detect light as well as heavy charged particles and handle high count rates. The whole array was made in several detachable segments in order that any part of the array can be used separately or in conjunction with other detector systems as and when need arises. Part of ChAKRA has already been successfully used in several experiments at low energy($\lesssim$ 10MeV/A. It is hoped that, with the advent of upcoming new accelerators  (superconducting cyclotron, rare ion beam facility) at VECC, the ChAKRA will be immensely useful for next generation nuclear reaction studies at VECC.
\section*{References}
\bibliography{samir}

\begin{thebibliography}{10}
\expandafter\ifx\csname url\endcsname\relax
  \def\url#1{\texttt{#1}}\fi
\expandafter\ifx\csname urlprefix\endcsname\relax\def\urlprefix{URL }\fi
\expandafter\ifx\csname href\endcsname\relax
  \def\href#1#2{#2} \def\path#1{#1}\fi

\bibitem{INDRA:NIMA:357:1995:418}
J.~Pouthas, B.~Borderie, R.~Dayras, E.~Plagnol, M.~Rivet, F.~Saint-Laurent,
  J.~Steckmeyer, G.~Auger, C.~Bacri, S.~Barbey, A.~Barbier, A.~Benkirane,
  J.~Benlliure, B.~Berthier, E.~Bougamont, P.~Bourgault, P.~Box, R.~Bzyl,
  B.~Cahan, Y.~Cassagnou, D.~Charlet, J.~Charvet, A.~Chbihi, T.~Clerc,
  N.~Copinet, D.~Cussol, M.~Engrand, J.~Gautier, Y.~Huguet, O.~Jouniaux,
  J.~Laville, P.~L. Botlan, A.~Leconte, R.~Legrain, P.~Lelong, M.~L. Guay,
  L.~Martina, C.~Mazur, P.~Mosrin, L.~Olivier, J.~Passerieux, S.~Pierre,
  B.~Piquet, E.~Plaige, E.~Pollacco, B.~Raine, A.~Richard, J.~Ropert,
  C.~Spitaels, L.~Stab, D.~Sznajderman, L.~Tassan-got, J.~Tillier, M.~Tripon,
  P.~Vallerand, C.~Volant, P.~Volkov, J.~Wieleczko, G.~Wittwer,
  \href{http://www.sciencedirect.com/science/article/pii/0168900294015430}{Indra,
  a 4pi charged product detection array at ganil}, Nuclear Instruments and
  Methods in Physics Research Section A: Accelerators, Spectrometers, Detectors
  and Associated Equipment 357~(2) (1995) 418 -- 442.
\newblock \href
  {http://dx.doi.org/https://doi.org/10.1016/0168-9002(94)01543-0}
  {\path{doi:https://doi.org/10.1016/0168-9002(94)01543-0}}.
\newline\urlprefix\url{http://www.sciencedirect.com/science/article/pii/0168900294015430}

\bibitem{HIRA:NIMA:583:2007:302}
M.~Wallace, M.~Famiano, M.-J. van Goethem, A.~Rogers, W.~Lynch, J.~Clifford,
  F.~Delaunay, J.~Lee, S.~Labostov, M.~Mocko, L.~Morris, A.~Moroni, B.~Nett,
  D.~Oostdyk, R.~Krishnasamy, M.~Tsang, R.~de~Souza, S.~Hudan, L.~Sobotka,
  R.~Charity, J.~Elson, G.~Engel,
  \href{http://www.sciencedirect.com/science/article/pii/S016890020701947X}{The
  high resolution array (hira) for rare isotope beam experiments}, Nuclear
  Instruments and Methods in Physics Research Section A: Accelerators,
  Spectrometers, Detectors and Associated Equipment 583~(2) (2007) 302 -- 312.
\newblock \href {http://dx.doi.org/https://doi.org/10.1016/j.nima.2007.08.248}
  {\path{doi:https://doi.org/10.1016/j.nima.2007.08.248}}.
\newline\urlprefix\url{http://www.sciencedirect.com/science/article/pii/S016890020701947X}

\bibitem{CHIMERA:1995}
S.~Aiello, A.~Anzalone, Baldo, M.~Cardella, G.~Cavallaroo, E.~Filippo,
  S.~Pietro, A.Di ans~Feminoc, P.~Figuera, P.~Guazzonid, C.~Iacono-Manno,
  G.~Lanzanob, U.~Lombardo, S.~Nigro, A.~Muumarra, A.~Pagano, M.~Papab,
  S.~Pirroneb, G.~Politi, F.~Porto, A.~Rapisardab, F.~Rizzo, S.~Sambatarob,
  M.~Sperduto, C.~Suterah, L.~Zettad, {Chimera: a project of a 4$\pi$ detector
  for heavy ion reactions studies at intermediate energy}, Nucl. Inst. Meth.
  AS83 ((1995)) 461.

\bibitem{LASSA:473:2001:302}
B.~Davin, R.~de~Souza, R.~Yanez, Y.~Larochelle, R.~Alfaro, H.~Xu, A.~Alexander,
  K.~Bastin, L.~Beaulieu, J.~Dorsett, G.~Fleener, L.~Gelovani, T.~Lefort,
  J.~Poehlman, R.~Charity, L.~Sobotka, J.~Elson, A.~Wagner, T.~Liu, X.~Liu,
  W.~Lynch, L.~Morris, R.~Shomin, W.~Tan, M.~Tsang, G.~Verde, J.~Yurkon,
  \href{http://www.sciencedirect.com/science/article/pii/S0168900201002959}{Lassa:
  a large area silicon strip array for isotopic identification of charged
  particles}, Nuclear Instruments and Methods in Physics Research Section A:
  Accelerators, Spectrometers, Detectors and Associated Equipment 473~(3)
  (2001) 302 -- 318.
\newblock \href
  {http://dx.doi.org/https://doi.org/10.1016/S0168-9002(01)00295-9}
  {\path{doi:https://doi.org/10.1016/S0168-9002(01)00295-9}}.
\newline\urlprefix\url{http://www.sciencedirect.com/science/article/pii/S0168900201002959}

\bibitem{Fazia:EPJA:50:2014:47}
{The FAZIA Collaboration}, R.~Bougault, G.~Poggi, S.~Barlini, B.~Borderie,
  G.~Casini, A.~Chbihi, N.~Le~Neindre, M.~P{\^a}rlog, G.~Pasquali,
  S.~Piantelli, Z.~Sosin, G.~Ademard, R.~Alba, A.~Anastasio, S.~Barbey,
  L.~Bardelli, M.~Bini, A.~Boiano, M.~Boisjoli, E.~Bonnet, R.~Borcea,
  B.~Bougard, G.~Brulin, M.~Bruno, S.~Carboni, C.~Cassese, F.~Cassese,
  M.~Cinausero, L.~Ciolacu, I.~Cruceru, M.~Cruceru, B.~D'Aquino, B.~De~Fazio,
  M.~Degerlier, P.~Desrues, P.~Di~Meo, J.~A. Due{\~{n}}as, P.~Edelbruck,
  S.~Energico, M.~Falorsi, J.~D. Frankland, E.~Galichet, K.~Gasior,
  F.~Gramegna, R.~Giordano, D.~Gruyer, A.~Grzeszczuk, M.~Guerzoni, H.~Hamrita,
  C.~Huss, M.~Kajetanowicz, K.~Korcyl, A.~Kordyasz, T.~Kozik, P.~Kulig,
  L.~Lavergne, E.~Legou{\'e}e, O.~Lopez, J.~{\L}ukasik, C.~Maiolino, T.~Marchi,
  P.~Marini, I.~Martel, V.~Masone, A.~Meoli, Y.~Merrer, L.~Morelli, F.~Negoita,
  A.~Olmi, A.~Ordine, G.~Paduano, C.~Pain, M.~Pa{\l}ka, G.~Passeggio,
  G.~Pastore, P.~Paw{\l}owski, M.~Petcu, H.~Petrascu, E.~Piasecki,
  G.~Pontoriere, E.~Rauly, M.~F. Rivet, R.~Rocco, E.~Rosato, L.~Roscilli,
  E.~Scarlini, F.~Salomon, D.~Santonocito, V.~Seredov, S.~Serra, D.~Sierpowski,
  G.~Spadaccini, C.~Spitaels, A.~A. Stefanini, G.~Tobia, G.~Tortone,
  T.~Twar{\'o}g, S.~Valdr{\'e}, A.~Vanzanella, E.~Vanzanella, E.~Vient,
  M.~Vigilante, G.~Vitiello, E.~Wanlin, A.~Wieloch, W.~Zipper,
  \href{https://doi.org/10.1140/epja/i2014-14047-4}{The fazia project in
  europe: R{\&}d phase}, The European Physical Journal A 50~(2) (2014) 47.
\newblock \href {http://dx.doi.org/10.1140/epja/i2014-14047-4}
  {\path{doi:10.1140/epja/i2014-14047-4}}.
\newline\urlprefix\url{https://doi.org/10.1140/epja/i2014-14047-4}

\bibitem{micron}
\url{http://www.micronsemiconductor.co.uk}.

\bibitem{Scionix}
\url{http://www.scionix.nl/}.

\bibitem{LISE}
\textsc{LISE} ++, \url{http://lise.nscl.msu.edu/lise.html}.

\bibitem{GUINET:NIMA:278:1989:614}
D.~Guinet, B.~Chambon, B.~Cheynis, A.~Demeyer, D.~Drain, X.~Hu, C.~Pastor,
  L.~Vagneron, K.~Zaid, A.~Giorni, D.~Heuer, A.~Lleres, J.~Viano,
  \href{http://www.sciencedirect.com/science/article/pii/0168900289908899}{Using
  the combination csl(ti) and photodiode for identification and energy
  measurement of light particles}, Nuclear Instruments and Methods in Physics
  Research Section A: Accelerators, Spectrometers, Detectors and Associated
  Equipment 278~(2) (1989) 614 -- 616.
\newblock \href
  {http://dx.doi.org/https://doi.org/10.1016/0168-9002(89)90889-9}
  {\path{doi:https://doi.org/10.1016/0168-9002(89)90889-9}}.
\newline\urlprefix\url{http://www.sciencedirect.com/science/article/pii/0168900289908899}

\bibitem{saint-gobain}
\url{https://www.crystals.saint-gobain.com/}.

\bibitem{mesytec}
\url{https://www.mesytec.com}.

\bibitem{DAE:63:2018:678}
S.~Manna, C.~Bhattacharya, T.~Rana, S.~Kundu, R.~Pandey, A.~Sen, D.~Paul,
  P.~Roy, T.~Ghosh, G.~Mukherjee, S.~Mukhopadhyay, S.~Nandi, J.~Meena, R.~Saha,
  A.~Saha, J.~Sahoo, S.~Dalal, \href{www.sympnp.org/proceedings}{Cluster
  emission studies in 28,29si*}, DAE-BRNS Symp 63 (2018) 678--679.
\newline\urlprefix\url{www.sympnp.org/proceedings}

\bibitem{Tapan:PRC:78:2008:027602}
T.~K. Rana, C.~Bhattacharya, S.~Kundu, K.~Banerjee, S.~Bhattacharya, A.~Dey,
  T.~K. Ghosh, G.~Mukherjee, J.~K. Meena, D.~Gupta, S.~Mukhopadhyay, D.~Pandit,
  S.~R. Banerjee, A.~Roy, P.~Dhara,
  \href{https://link.aps.org/doi/10.1103/PhysRevC.78.027602}{Consistency of
  nuclear thermometric measurements at moderate excitation}, Phys. Rev. C 78
  (2008) 027602.
\newblock \href {http://dx.doi.org/10.1103/PhysRevC.78.027602}
  {\path{doi:10.1103/PhysRevC.78.027602}}.
\newline\urlprefix\url{https://link.aps.org/doi/10.1103/PhysRevC.78.027602}

\bibitem{ROOT}
\url{https://root.cern.ch/}.

\bibitem{DAE:57:2012:864}
S.~Kundu, C.~Bhattacharya, T.~K. Rana, K.~Banerjee, S.~Bhattacharya, J.~K.
  Meena, R.~Saha, G.~Mukherjee, T.~K. Ghosh, R.~Pandey, P.~Roy, M.~Gohil,
  V.~Srivastava, A.~Dey, G.~Pal, S.~Roy, S.~R. Bajirao, C.~Nandi,
  \href{www.sympnp.org/proceedings}{Charged particle detector array:
  45$^{0}$-175$^{0}$*}, DAE-BRNS Symp 57 (2012) 864--865.
\newline\urlprefix\url{www.sympnp.org/proceedings}

\bibitem{DAE:51:2006:618}
S.~Kundu, K.~Banerjee, T.~Rana, C.~Bhattacharya, P.~C. Rout, A.~Mitra,
  E.~Mirgule, V.~Nanal, S.~Kumar, V.~Datar, D.~Chakrabarty, S.~Bhattacharya,
  A.~Dey, G.~Mukherjee, T.~Ghosh, D.~Gupta, J.~Meena,
  \href{http://www.sympnp.org/}{Testing of a csi (tl) detector and its energy
  calibration}, DAE-BRNS Symp 51 (2006) 618--619.
\newline\urlprefix\url{http://www.sympnp.org/}

\bibitem{samir:JoP:390-2012-012075}
S.~Kundu, S.~Bhattacharya, J.~K. Meena, T.~K. Ghosh, T.~Bhattacharjee,
  P.~Mukhopadhyay, C.~Bhattacharya, T.~K. Rana, K.~Banerjee, G.~Mukherjee,
  S.~R. Banerjee, D.~L. Bandyopadhyay, M.~Ahammed, P.~Bhattacharya,
  \href{http://stacks.iop.org/1742-6596/390/i=1/a=012075}{A large high vacuum
  reaction chamber for nuclear physics research at vecc, kolkata}, Journal of
  Physics: Conference Series 390~(1) (2012) 012075.
\newline\urlprefix\url{http://stacks.iop.org/1742-6596/390/i=1/a=012075}

\bibitem{Tapan:88:2013:021601}
T.~K. Rana, S.~Bhattacharya, C.~Bhattacharya, S.~Kundu, K.~Banerjee, T.~K.
  Ghosh, G.~Mukherjee, R.~Pandey, P.~Roy, V.~Srivastava, M.~Gohil, J.~K. Meena,
  H.~Pai, A.~K. Saha, J.~K. Sahoo, R.~M. Saha,
  \href{https://link.aps.org/doi/10.1103/PhysRevC.88.021601}{Estimation of
  direct components of the decay of the hoyle state}, Phys. Rev. C 88 (2013)
  021601.
\newblock \href {http://dx.doi.org/10.1103/PhysRevC.88.021601}
  {\path{doi:10.1103/PhysRevC.88.021601}}.
\newline\urlprefix\url{https://link.aps.org/doi/10.1103/PhysRevC.88.021601}

\bibitem{Vishal:PRC:91:2015:54611}
V.~Srivastava, C.~Bhattacharya, T.~K. Rana, S.~Manna, S.~Kundu,
  S.~Bhattacharya, K.~Banerjee, P.~Roy, R.~Pandey, G.~Mukherjee, T.~K. Ghosh,
  J.~K. Meena, T.~Roy, A.~Chaudhuri, M.~Sinha, A.~Saha, M.~A. Asgar, A.~Dey,
  S.~Roy, M.~M. Shaikh,
  \href{https://link.aps.org/doi/10.1103/PhysRevC.91.054611}{Experimental study
  of $^{26}\mathrm{Al}$ through the $1n$ pick-up reaction
  $^{27}\mathrm{Al}(d,t)$}, Phys. Rev. C 91 (2015) 054611.
\newblock \href {http://dx.doi.org/10.1103/PhysRevC.91.054611}
  {\path{doi:10.1103/PhysRevC.91.054611}}.
\newline\urlprefix\url{https://link.aps.org/doi/10.1103/PhysRevC.91.054611}

\bibitem{Vishal:93:2016:044601}
V.~Srivastava, C.~Bhattacharya, T.~K. Rana, S.~Manna, S.~Kundu,
  S.~Bhattacharya, K.~Banerjee, P.~Roy, R.~Pandey, G.~Mukherjee, T.~K. Ghosh,
  J.~K. Meena, T.~Roy, A.~Chaudhuri, M.~Sinha, A.~K. Saha, M.~A. Asgar, A.~Dey,
  S.~Roy, M.~M. Shaikh,
  \href{https://link.aps.org/doi/10.1103/PhysRevC.93.044601}{Experimental
  investigation of $t=1$ analog states of $^{26}\mathrm{Al}$ and
  $^{26}\mathrm{Mg}$}, Phys. Rev. C 93 (2016) 044601.
\newblock \href {http://dx.doi.org/10.1103/PhysRevC.93.044601}
  {\path{doi:10.1103/PhysRevC.93.044601}}.
\newline\urlprefix\url{https://link.aps.org/doi/10.1103/PhysRevC.93.044601}

\bibitem{Santu:PRC:94:2016:051601}
S.~Manna, T.~K. Rana, C.~Bhattacharya, S.~Bhattacharya, S.~Kundu, K.~Banerjee,
  P.~Roy, R.~Pandey, V.~Srivastava, A.~Chaudhuri, T.~Roy, T.~K. Ghosh,
  G.~Mukherjee, J.~K. Meena, S.~K. Pandit, K.~Mahata, A.~Shrivastava, V.~Nanal,
  \href{https://link.aps.org/doi/10.1103/PhysRevC.94.051601}{Survival of
  cluster correlation in dissipative binary breakup of
  $^{24,25}\mathrm{Mg}^{*}$}, Phys. Rev. C 94 (2016) 051601.
\newblock \href {http://dx.doi.org/10.1103/PhysRevC.94.051601}
  {\path{doi:10.1103/PhysRevC.94.051601}}.
\newline\urlprefix\url{https://link.aps.org/doi/10.1103/PhysRevC.94.051601}

\bibitem{ratnesh:95:2017:064603}
R.~Pandey, S.~Kundu, C.~Bhattacharya, K.~Banerjee, T.~K. Rana, S.~Manna,
  G.~Mukherjee, J.~K. Meena, A.~Chaudhuri, T.~Roy, P.~Roy, M.~A. Asgar,
  V.~Srivastava, A.~Dey, M.~Sinha, T.~K. Ghosh, S.~Bhattacharya, S.~K. Pandit,
  K.~Mahata, P.~Patle, S.~Pal, A.~Shrivastava, V.~Nanal,
  \href{https://link.aps.org/doi/10.1103/PhysRevC.95.064603}{Fragment emission
  mechanism in the $^{32}\mathrm{S}+^{12}\mathrm{C}$ reaction}, Phys. Rev. C 95
  (2017) 064603.
\newblock \href {http://dx.doi.org/10.1103/PhysRevC.95.064603}
  {\path{doi:10.1103/PhysRevC.95.064603}}.
\newline\urlprefix\url{https://link.aps.org/doi/10.1103/PhysRevC.95.064603}

\end{thebibliography}
\end{document}